\documentstyle[sprocl,epsf,epsfig]{article}

\def\PRL{\em Phys. Rev. Lett.}

\def\be{\begin{equation}}
\def\ee{\end{equation}}
\def\bea{\begin{eqnarray}}
\def\eea{\end{eqnarray}}

\def\beq{\begin{equation}}
\def\eeq{\end{equation}}
\def\bea{\begin{eqnarray}}
\def\eea{\end{eqnarray}}
\def\bq{\begin{quote}}
\def\eq{\end{quote}}
\def\PL{{ \it Phys. Lett.} }
\def\PRL{{\it Phys. Rev. Lett.} }
\def\NP{{\it Nucl. Phys.} }
\def\PR{{\it Phys. Rev.} }

\def\MPL{{\it Mod. Phys. Lett.} }

\def\gappeq{\mathrel{\rlap {\raise.5ex\hbox{$>$}}
{\lower.5ex\hbox{$\sim$}}}}
 
\def\lappeq{\mathrel{\rlap{\raise.5ex\hbox{$<$}}
{\lower.5ex\hbox{$\sim$}}}}

\begin{document}

\pagestyle{empty}
\begin{flushright}
{CERN-TH/96-139}
\end{flushright}
\vspace*{5mm}
\begin{center}
{\bf STRING COSMOLOGY AND RELIC GRAVITATIONAL } \\ 
{\bf RADIATION} \\  
\vspace*{1cm} {\bf G. Veneziano} \\
\vspace*{0.5cm}
Theoretical Physics Division, CERN \\
CH - 1211 Geneva 23 \\ 
\vspace*{2cm}  
{\bf ABSTRACT} \\ \end{center}
\vspace*{5mm}
\noindent
String theory counterparts to Einstein's gravity, cosmology and inflation are
described. A very tight upper
bound on the Cosmic Gravitational Radiation Background (CGRB) 
of standard  inflation is  shown to be evaded in string cosmology, 
while an interesting  signal in the phenomenologically interesting frequency
range is all but excluded.  The generic features of such a stringy CGRB are
 presented.

\vspace*{3cm}
\noindent
\rule[.1in]{12cm}{.002in}

\noindent
$^{1)}$ Talk presented at the International Conference on Gravitational Waves:
Sources and Detectors,  Cascina (Pisa), March 1996.

\begin{flushleft} CERN-TH/96-139 \\
June 1996
\end{flushleft}
\vfill\eject

\setcounter{page}{1}
\pagestyle{plain}

\title{STRING COSMOLOGY AND RELIC GRAVITATIONAL RADIATION}

\author{ G. VENEZIANO }

\address{Theoretical Physics Division, CERN \\
CH - 1211 Geneva 23}


\maketitle\abstracts{
String theory counterparts to Einstein's gravity, cosmology and inflation are
described. A very tight upper
bound on the Cosmic Gravitational Radiation Background (CGRB) 
of standard  inflation is  shown to be evaded in string cosmology, 
while an interesting  signal in the phenomenologically interesting frequency
range is all but excluded.  The generic features of such a stringy
CGRB are presented.
}

\section{Introduction}

 In this talk I will first  explain why string theory offers an interesting
alternative to Einstein's gravity and cosmology. The standard  post-big-bang
picture emerges as just the {\it late-time} history of a Universe which, in a
prehistoric  ({\it pre-big-bang}) era, underwent an inflationary
expansion driven by the growth of the universal coupling of the theory.

I will then turn to describing one of the most
interesting physical consequences of this new scenario: the production of a
Cosmic Gravitational Radiation Background (CGRB), which could by far exceed, 
in the relevant frequency range, the one predicted by
ordinary inflationary models.

I will leave the detailed discussion of the near-future prospects
for observability of our CGRB to the following talk by R. Brustein and refer
you, for more details on the scenario and the computations, to the collection of
papers on string cosmology appearing on WWW under:

http:/www.to.infn.it/teorici/gasperini/

The precious collaboration of Maurizio Gasperini throughout the development of
the pre-big-bang scenario, and the additional one of Ramy Brustein, Massimo
Giovannini and Slava Mukhanov in working out its  consequences for gravitational
perturbations, are gratefully acknowledged.
 
\section{Einstein Gravity and Standard Cosmology}
 
In order to introduce string gravity and a cosmological model
based on it, I will first recall a few known facts about Einstein gravity
and standard cosmology (see for instance \cite{W}).

The well-known Einstein equations:
\beq
R_{\mu \nu} - 1/2 g_{\mu \nu} R + \Lambda g_{\mu \nu} = - 8 \pi G T_{\mu \nu} 
\eeq
follow from setting to zero the variation of the Einstein--Hilbert action
(I will use $c= \hbar =1$ throughout):
\beq
S = - {1 \over 16 \pi G} \int d^4x \sqrt{-g}~
\left[ R - 2 \Lambda \right]  + S_{matter}\; .
\eeq

Einstein cosmology  follows from Einstein's equations upon
insertion of a homogeneous (and, for simplicity, spatially flat) 
ansatz for the
metric: 
\beq
ds^2 \equiv g_{\mu \nu} dx^{\mu}dx^{\nu} = dt^2 - a(t)^2 dx^i dx^i
\eeq
and after assuming that also matter is homogeneously distributed.
The Einstein--Friedman equations (of which only two are independent) then
follow: 
\bea
H^2 \equiv  (\dot {a} /a )^2 &=& {8 \pi G \over 3} \rho + {\Lambda \over 3}
\nonumber \\ 
\dot{H} + H^2 \equiv (\ddot a /a) &=& - { 4 \pi G \over 3} (\rho + 3 p)
+ {\Lambda \over 3} \nonumber \\
 \dot{\rho} &=& -3 H (\rho + p)
\label{EF}
\eea
where the matter energy density $\rho$ and  pressure $p$ are defined in terms
of $T_{\mu \nu}$ by $T_{\mu}^{\nu} = {\rm diag}~(\rho, -p , -p , -p)$.
Notice that the effect of a non-vanishing cosmological constant $\Lambda$
is equivalent to that of a special kind of matter, the ``vacuum", with
$\rho_{vac} = - p_{vac} = {\Lambda \over 8 \pi G}$. Normal matter has $(\rho +
3p)>0$ and therefore leads to a decelerated expansion of the Universe: in
particular, a matter-dominated Universe ($p/\rho \sim 0$) expands like
$t^{2/3}$, while a radiation-dominated Universe ($p/\rho \sim 1/3$) expands like
$t^{1/2}$.

A trivial but important remark for the following discussion:
if we regard (as we should) the first of eqs. (\ref{EF}) as expressing the
vanishing of the total energy of the matter-plus-gravity system, we see that the
expansion of the Universe contributes with a ${\it negative}$ kinetic energy to
such an equation.

Inflation, i.e. a long phase of accelerated expansion of
the Universe ($\dot a , \ddot a > 0$), is badly needed in order to solve the
outstanding problems of the standard cosmological model \cite{INFL}.
Unlike ordinary matter, a cosmological constant can easily do the job. 
The same is
true of potential energy originating from a scalar field 
(the so-called inflaton)
which, during some cosmic epoch, was approximately frozen  away from the
 minimum of its potential and  thus provided an effective (positive)
cosmological constant $\Lambda_{eff} = 8 \pi G V$. In this case a (quasi) de
Sitter exponential expansion of the Universe takes place:
\beq
 a(t) \sim \exp (H t) \; , \; ~ H^2 =  {8 \pi \over 3} G V = \Lambda_{eff}/3 ~ .
\eeq

\section{  The disappointing CGRB of standard inflation}

In standard potential-energy-driven inflation, while the inflaton  slowly
rolls down to the true minimum of the potential (where, by assumption, the 
potential energy is very small), 
the Hubble parameter $H$ stays constant or decreases slowly. The Hubble radius
$H^{-1}$ thus remains constant (or increases slowly) during the inflationary
epoch and then starts to grow like cosmic time $t$  during the radiation-
and matter-dominated eras.

In Fig. 1 this behaviour of the Hubble radius is plotted together with
the behaviour  of different physical scales which, by definition, grow like the
scale factor $a(t)$ itself. It is easily found that scales cross the ``horizon"
outward (exit) during  inflation and cross it again inward (re-enter) during the
matter- or radiation-dominated epochs. Larger scales exit earlier and re-enter
later than shorter scales. In order to solve the homogeneity problem of standard
cosmology, it is necessary that the scale corresponding to the present horizon,
$O(H_0^{-1})$, once upon a time, was inside the horizon. For this to happen
a total red-shift of about 
 \beq 
z_{infl} \equiv {a_{end} \over a_{beg}} >
10^{30} 
\eeq
during inflation is needed \cite{INFL}.

\begin{figure}
\hglue1cm
\epsfig{figure=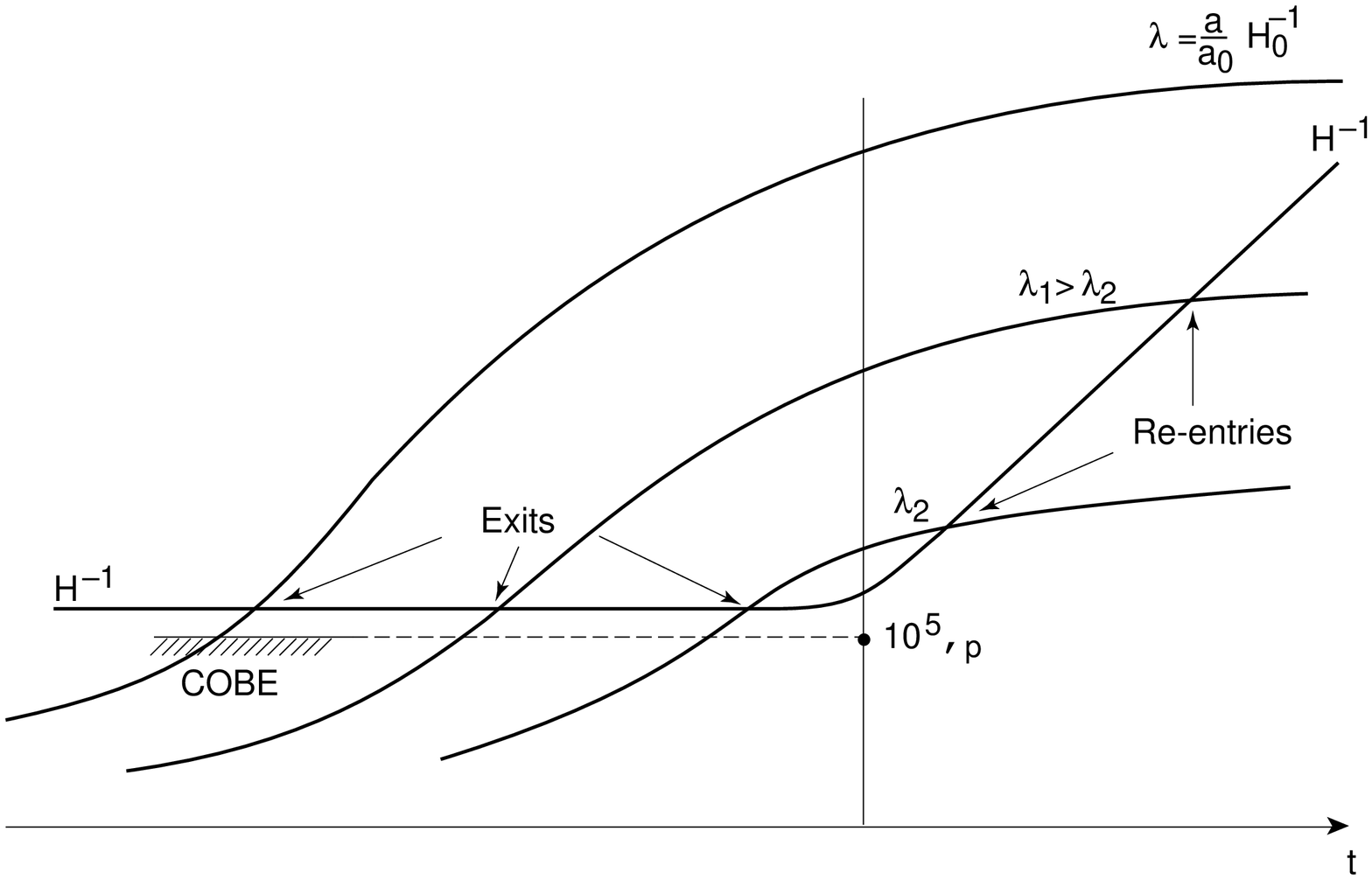,height=2.5in}
\begin{center} Figure 1 \end{center}
\end{figure}

One of the celebrated bonuses of inflation \cite{INFL}
 is a natural explanation of the
origin of large-scale structure. Let us assume that, initially, there were no
inhomogeneities other than the minimal ones due to quantum mechanics. In other
words, let us ask ourselves whether the  origin of a structure in the Universe can
be found in the initial vacuum quantum fluctuations. Vacuum fluctuations of the
metric (defined as usual by $g_{\mu \nu} = \eta_{\mu \nu} + h_{\mu \nu}$)
with wavelength $\lambda$ have a typical magnitude
\beq
\delta h (\lambda) \sim {\ell_P \over \lambda}\; ,
\label{VF}
\eeq
where $\ell_P = \sqrt{G}$ is the Planck length whose magnitude controls the
size of quantum gravity effects.
We see that the shorter the wavelength the larger the quantum fluctuation.

It is not hard to show that these original perturbations 
(inhomogeneities) are adiabatically damped (i.e. they follow eq. (\ref{VF}) with 
$\lambda \sim a$) as long
as their physical wavelength stays inside the horizon, while they
freeze-out (stay constant) after going outside. Since larger wavelengths spend
a longer time outside the horizon (Fig.1) they stay frozen for a longer time. In
other words there is a competition between two effects: quantum mechanics
favours short scales while  classical freeze-out favours large scales. 

Combining the two effects leads to a
simple and suggestive formula \cite{INFL} for the present magnitude of tensor
metric perturbations i.e of  gravitational waves (GW):
\beq
\rho_{\gamma}^{-1} {d \rho \over d \log \omega} \equiv {\Omega_{GW}(\omega)
 \over
\Omega_{\gamma}} \sim (\ell_P H)^2|_{ex} \; ,
\label{OmegaGW}
\eeq
where $\Omega_{\gamma} = {\rho_{\gamma} \over \rho_{cr}} \sim 10^{-4}$ and
$\Omega_{GW}(\omega) = \rho_{cr}^{-1} {d \rho_{GW} \over d \log \omega}$   
and the label $ex$ indicates that $l_P H$ has to be
evaluated, for each scale $\lambda$, at the time of its exit.
This is the crucial quantity for the GW yield at any given
frequency.  As we have explained, $H$ is constant
or slowly decreasing during inflation, hence the same is true of the GW spectrum
as a function of $\omega$. This is the celebrated (quasi) scale-invariant
Harrison--Zeldovich spectrum, which appears to be quite efficient for generating
the observed large-scale structure (if combined with an appropriate model for
dark matter).

Unfortunately, for the purpose of this talk, the above result is bad news, i.e.
represents a disappointing spectrum of GW in the relevant frequency region.
Indeed, COBE's observation \cite{CO} of a ${\Delta T \over T}$ of order
$10^{-5}$ at large angular scales implies $ H^{-1} > 10^5 \ell_P$ when scales of the
order of the present Hubble radius went out of the horizon, and  an even
smaller value when shorter scales did (see again Fig.1). Inserting such a limit in
eq. (\ref{OmegaGW}), we  immediately arrive at:
 \beq
\Omega_{GW}(\omega) < 10^{-14}~ {\rm to} ~ 10^{-15}
\label{bound}
\eeq
in the interesting (Hz to MHz) frequency range. This upper limit makes the CGRB
produced by ordinary inflation an unobservable signal for some time to come $\dots$

\section{String Gravity}

Being a theory of extended objects, string theory contains a
fundamental length scale
 $\lambda_s$, a built-in ultraviolet
(short-distance) cut-off \cite{VE}. As a result,
string gravity differs from Einstein gravity in a subtle and essential way.
Instead of the action (1.1), string theory gives \cite{EFF}:
\bea
\Gamma_{eff} &=& \frac{1}{2} \int d^4x \sqrt{-g}~ e^{-\phi}
\left[\lambda^{-2}_s ( R
+ \partial_\mu \phi \partial^\mu\phi) + F^2_{\mu\nu} + \bar \psi
D\llap{$/$} \psi \right] \nonumber \\ && + \left[ {\rm higher~orders~in}~
\lambda_s^2 \cdot \partial^2 \right]     + \left[ {\rm
higher~orders~in}~e^\phi \right]~. 
\label{31}
\eea

 As indicated in (\ref{31}), string gravity has (actually needs!) a new
particle/field,
the so-called dilaton $\phi$, a scalar particle.  It enters
 $\Gamma_{eff}$ as a Jordan--Brans--Dicke \cite{JBD}
 scalar with a ``small" negative
 $\omega_{BD}$ parameter, $\omega_{BD} = -1$.
Bounds on the present rate of variation of $\alpha$ and $G$ imply
that, today, $\dot {\phi} < H_0$, while 
 precision tests \cite{EP} of the equivalence principle put an upper (lower) 
 limit on the range of the dilaton-exchange force (on the dilaton mass)
\cite{TV}:
 \beq
 m_\phi > 10^{-4}~{\rm eV}~.
\label{34}
\eeq

Both problems are solved by assuming that a non-perturbative dilaton
potential has to be added to (\ref{31}). Such a potential will freeze
the dilaton
to its present value $\phi_0$  and make us recover Einstein's theory
(and its experimental successes) at late times.
  
The value $\phi_0$ provides \cite{Wi} today's unified value of 
the gauge  and  gravitational couplings
at energy scales of $O(\lambda_s^{-1})$. In formulae:
\bea
\ell^2_P & \equiv & 8 \pi G_N  = e^{\phi_0} \lambda^2_s ~, \nonumber \\
\alpha_{GUT}(\lambda^{-1}_s) &\simeq& \frac{e^{\phi_0}}{4\pi}~ = \frac{\ell_P^2}
{\lambda^2_s} ,
\label{32}
\eea
implying (from $\alpha_{GUT} \approx 1/20$) that the string-length
parameter
$\lambda_s$ is about $10^{-32}$~cm.  Note, however, that the above formulae,
in a
cosmological context in which $\phi$ evolves in time,
 can only be taken as  giving the
$\it{present}$ values of $\alpha$ and $\ell_P/\lambda_s$.  In the
scenario we
will advocate, both quantities were much smaller  in the very
early Universe!

Equation (\ref{31}) contains two dimensionless expansion parameters.
One of them, the above-mentioned $g^2 \equiv e^\phi$, controls the analogue of
   loop corrections in quantum field theory (QFT), while the other,
 $\lambda^2 \equiv \lambda_s^2 \cdot \partial^2$, controls
  string-size effects, which are of course absent from QFT.
Obviously, the expansion in $\lambda^2$ is reliable at small curvatures 
(derivatives), i.e. at energies smaller than the  string scale $\lambda_s^{-1}$,
while higher orders in $g^2$ will be  negligible at weak coupling.

The first and main assumption of our scenario is that the Universe started
its evolution in a regime that was perturbative with respect to
both expansions, i.e. in a region of weak coupling and small curvatures
(derivatives). During that phase the string-gravity equations take
the simple form:
\bea
R_{\mu \nu} + \nabla_{\mu} \nabla_{\nu} \phi = - \lambda_s^2~ e^{\phi}
T_{\mu \nu} \nonumber \\ 
R -  \nabla_{\mu} \phi \nabla^{\mu} \phi + 2 \nabla^2 \phi + 2 \Lambda = 0
\; ,    \eea
which are similar to Einstein's equations, yet substantially different.
As already stressed, we wish to recover general relativity at late
times; nonetheless, we want to take advantage of the difference for the
prehistory of the Universe.

Before closing this section I would like to briefly comment  on a point
that appears to be the source of much confusion: it is the
dilemma between working in the so-called string frame and working in
the more conventional Einstein frame. The two frames are not  to be confused
with different coordinate systems: they
are instead related by a local field redefinition, a conformal, dilaton-dependent
rescaling of the metric, to be precise.
 All physical quantities are independent
of the frame one is using. The question is: What should we call the metric?
Although, to a large extent, this is a question of taste, one's intuition may
work better with one definition than with another. Note also that, since the
dilaton is time-independent today, the two frames  now coincide.

Let us compare the virtues and problems inherent in each frame.
\begin{itemize}
\item[A)] {\bf String Frame.}
This is the metric appearing in the original ($\sigma$-model) action for the
string.
 Classical, weakly coupled strings sweep geodesic surfaces with respect
to this metric \cite{SV}. Also, the dilaton dependence of the 
low-energy effective action
takes the simple form indicated in (\ref{31}) only in the string frame. The
advantage of this frame is that the string cut-off is fixed and the same is true
of the value of the curvature at which higher orders in the $\sigma$-model
coupling  $\lambda$ become relevant. The main disadvantage is that the
gravitational action is not so easy to work with.

 \item[B)] {\bf Einstein Frame.} 
In this frame the pure gravitational action takes the standard
Einstein--Hilbert form. Consequently, this is the most convenient
frame for studying the cosmological evolution of metric
perturbations. The Planck length is fixed in this frame, while the
string length is dilaton- (hence generally time-) dependent. In the
Einstein frame, $\Gamma_{eff}$
 takes the form:
\bea
\Gamma_{eff} &=&  \int \frac{d^4x \sqrt{-g}}{16 \pi G_N}~ 
\left[ R + \partial_\mu \phi \partial^\mu \phi + e^{-\phi} F^2_{\mu\nu} +  
   \partial_\mu A \partial^\mu A + e^{\phi} m^2 A^2 \right]   \nonumber \\
&& + \left[ G_N e^{-\phi} R^2 + \dots \right] , 
\label{37}
\eea
 showing that the constancy of $G$ in this frame is only apparent, since 
masses are
dilaton-dependent (even at tree level). The same is true of the value of $R$ at
which higher order stringy corrections become important. 
\end{itemize}
For the above
reasons  I will choose to base the discussion (although not always the
calculations) in the string frame.

\section{String cosmology}

There is an exact (all-order) vacuum solution for (critical) superstring
theory. Unfortunately, it corresponds to a free theory ($g=0$ or $\phi =
-\infty$) in flat, ten-dimensional, Minkowski space-time, nothing like
the world we are living in today! Could this  instead have been the
{\it original} state of the Universe? 
 The very basic postulate of the pre-big-bang scenario \cite{SFD}, \cite{PBB}
 is that this is indeed the case.

Such a postulate is supported by the observation that, in the space of
homogeneous (and, for simplicity, spatially-flat) perturbative solutions to the
field equations,  the trivial vacuum is a very special,
$\it{unstable}$ solution. This is depicted in Fig.~2a for the simplest case of a
ten-dimensional cosmology in which three spatial dimensions evolve isotropically
while six ``internal" dimensions are static (it is easy to generalize
the discussion to the case of dynamical internal dimensions, but
then the picture becomes multidimensional).

\begin{figure}
\epsfig{figure=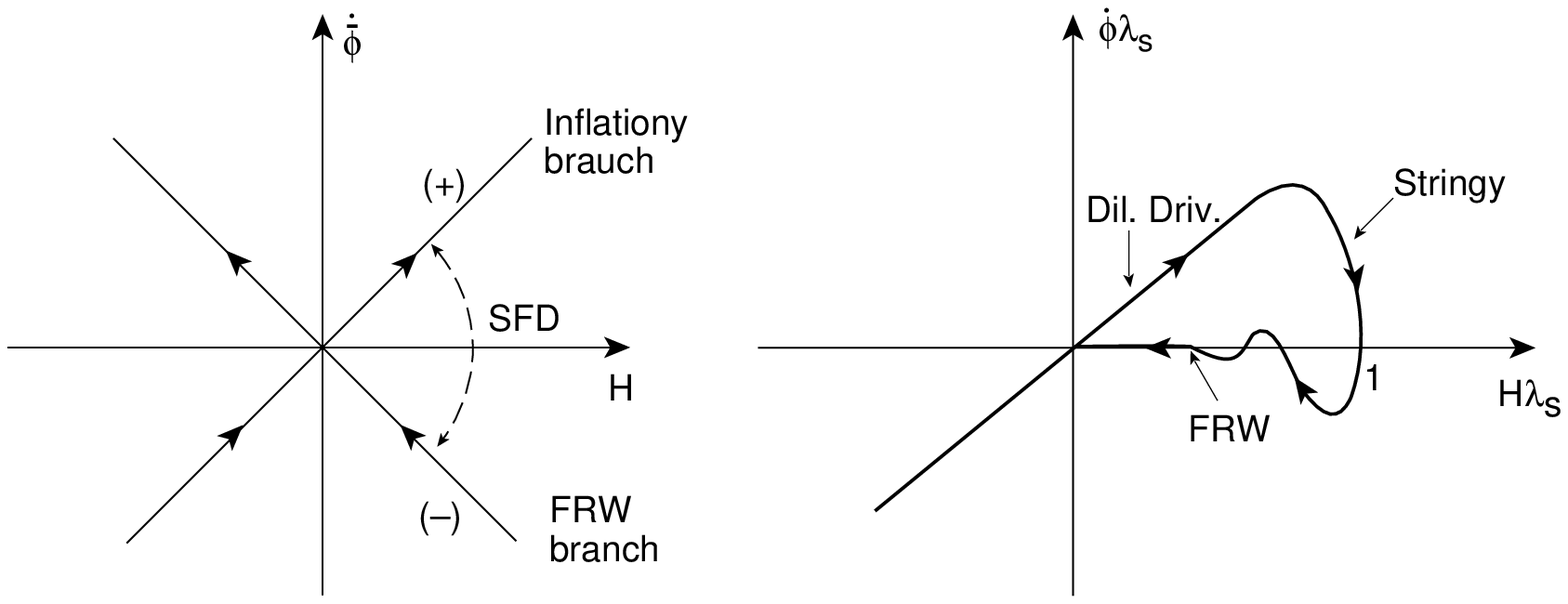,height=1.8in}
\begin{center} Figure 2 \end{center}
\end{figure}

The straight lines in the $H, \dot{\bar{\phi}}$ plane (where
$\dot{\bar{\phi}} \equiv \dot{\phi} - 3 H$) represent
the evolution of the scale factor and of the coupling constant
 as a function of the cosmic time parameter (arrows
along the lines show the direction of the time evolution).
As a consequence of a stringy symmetry,
known \cite{SFD}, \cite{rr} as Scale Factor Duality (SFD),
 there are two branches
(two straight lines). Furthermore, each branch  is
split by the origin in two time-reversal-related parts
(time reversal changes the sign of both
 $H$ and $\dot{\bar{\phi}}$).

As mentioned, the origin (the trivial vacuum) is an ``unstable"
 fixed point: a small perturbation in the direction of positive
$\dot{\bar{\phi}}$ makes the system evolve further and further
from the origin, meaning larger and larger coupling and
absolute value of the Hubble parameter.
This means an accelerated expansion (inflation) or an accelerated contraction.
 It is tempting to assume that those patches of the original
Universe that had the right kind of initial fluctuation  have  grown up
to become (by far) the largest fraction of the Universe today.
 
In order to arrive at a physically interesting scenario, however,
 we have to connect somehow the top-right inflationary branch
to  the bottom-right branch, since the latter is nothing
but the standard FRW cosmology, which has presumably prevailed
for the last few billion years or so.
Here the so-called {\it exit problem} of string cosmology arises.
 At lowest order
in $\lambda^2$ (small curvatures in string units)
the two branches do
not talk to each other. The inflationary (also called $+$) branch
has a singularity in the future (it takes a finite cosmic
time to reach $\infty$ in our gragh if one starts
from anywhere but the origin) while the FRW ($-$) branch
has a singularity in the past (the usual big-bang singularity).

It is widely believed that QST has a way to avoid the usual
singularities
of classical general relativity or at least a way to reinterpret
them \cite{KK}, \cite{top}. It thus looks reasonable to assume that the
inflationary
branch, instead of leading to a non-sensical singularity, will
evolve into the FRW branch at values of $\lambda^2$ of order unity.
This is schematically
shown in Fig. 2b, where we have gone back
from $\dot{\bar{\phi}}$ to $\dot{\phi}$
and we have implicitly taken into account the effects of a
non-vanishing dilaton potential at small $\phi$ in order to
freeze the dilaton at its present value.
The need for the branch change to occur at large $\lambda^2$,
first argued for in \cite{BV}, has  recently been proved \cite{KMO}.

\begin{figure}
\hglue1cm
\epsfig{figure=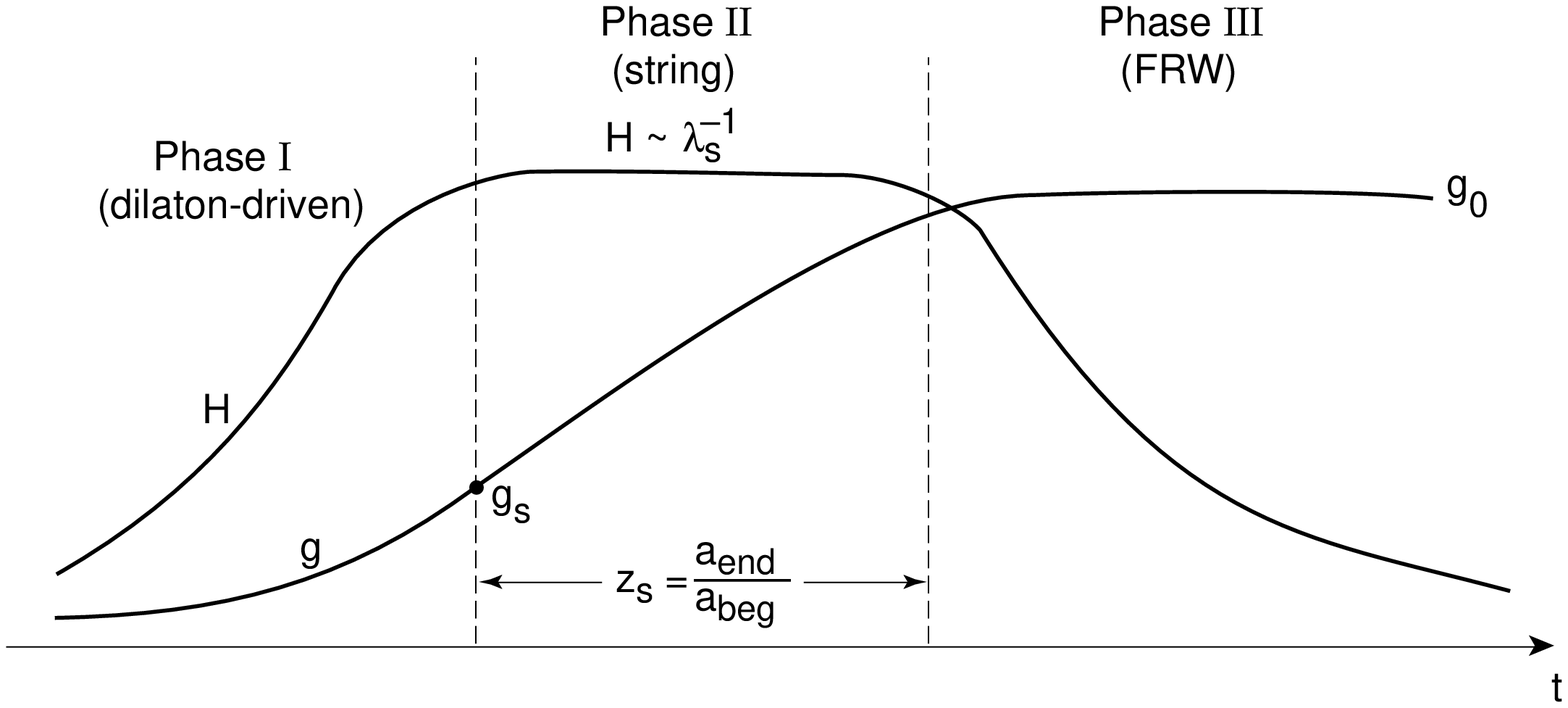,height=1.8in}
\begin{center} Figure 3 \end{center}
\end{figure}

There is a rather simple way to
parametrize a class of scenarios of the kind defined above.
They contain (roughly) three phases and two parameters, which can be
easily visualized in Fig. 3.

\begin{figure}
\hglue1cm
\epsfig{figure=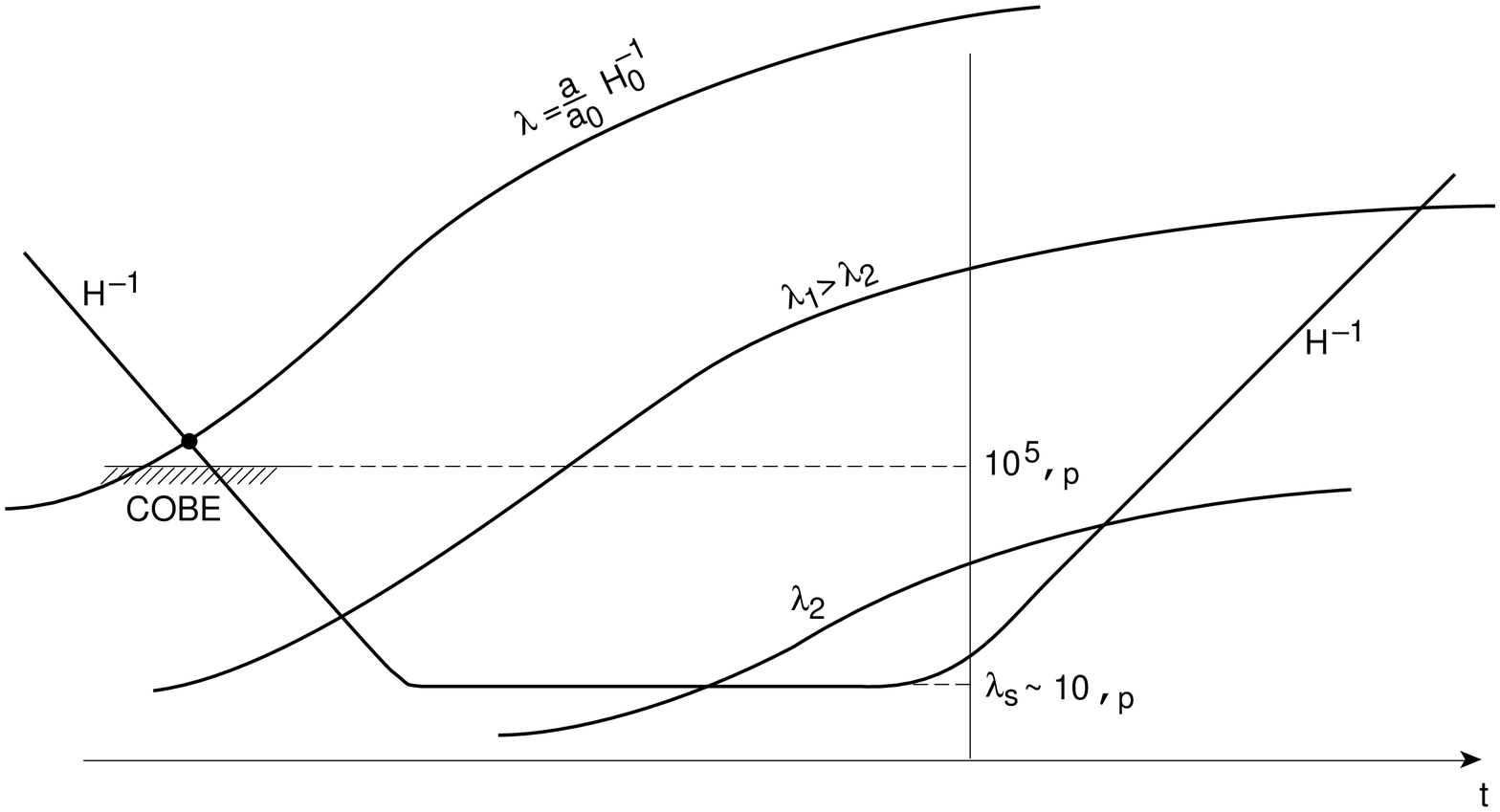,height=1.8in}
\begin{center} Figure 4 \end{center}
\end{figure}

In phase I the Universe evolves at $g^2, \lambda^2 \ll 1$
and  is thus close to the trivial vacuum. This phase
can  be studied using the tree-level low-energy effective action
(\ref{31}); it is characterized by a long period of dilaton-driven
inflation. The accelerated expansion of the Universe, instead of
originating from the potential
energy of an inflaton field, is
driven by the growth of the coupling constant (i.e. by the dilaton's
kinetic energy) 
 with $\dot{\phi} = 2\dot{g} / g \sim H$ during the whole phase.
Notice that, as for ordinary inflation, the negative value of the kinetic
energy associated with an expanding Universe is crucial. 
 
Phase I  supposedly  ends when the coupling $\lambda^2$
reaches values of $O(1)$, so that higher-derivative terms in
the effective action become relevant. Assuming that this happens
 while $g^2$ is still small (and thus the potential is
still negligible), the value $g_s$  of
$g$ at the end of phase I (the beginning
of phase II) is an arbitrary parameter (a modulus of the solution).
 
During phase II, the stringy version of the big bang,
 the curvature as well as $\dot\phi$ are
 assumed to remain fixed at their maximal value, given by
the string scale (i.e. we expect $\lambda \sim 1$).
The coupling $g$ will instead continue to grow from the value $g_s$
until, in turn, it
 reaches  values $O(1)$. At that point, thanks to a non-perturbative effect in
$g$, the string phase will come to an end and the
dilaton will be attracted to the true non-perturbative minimum of its
potential; the standard FRW cosmology can then start, provided
the Universe was by then heated-up and filled with radiation (see below).
The second important parameter of this scenario is the duration of phase
II or better the total red-shift, $z_s \equiv a_{end}/a_{beg}$,
which has occurred from the beginning to the end of the stringy phase.

\section{Perturbations and  CGRB in String Cosmology}

Starting from Fig. 3 let us draw the analogue of Fig. 1 for string cosmology,
i.e. the behaviour of the horizon and the way different scales
cross it. This is depicted in Fig. 4. Recalling eq. (\ref{OmegaGW})
we can easily understand why the bound (\ref{bound}) is now easily avoided.
All we need in order to satisfy COBE's constraint is that $\ell_P H$
had been small enough at the time when the present horizon's scale
crossed the Hubble radius during inflation (see hatched region in Fig. 4). Since
the horizon is shrinking during superinflation, this does 
not prevent $\ell_P H$ from
having been much smaller when scales of interest for GW detection (say above
$1$ Hz) crossed the horizon.

\begin{figure}
\hglue1cm
\epsfig{figure=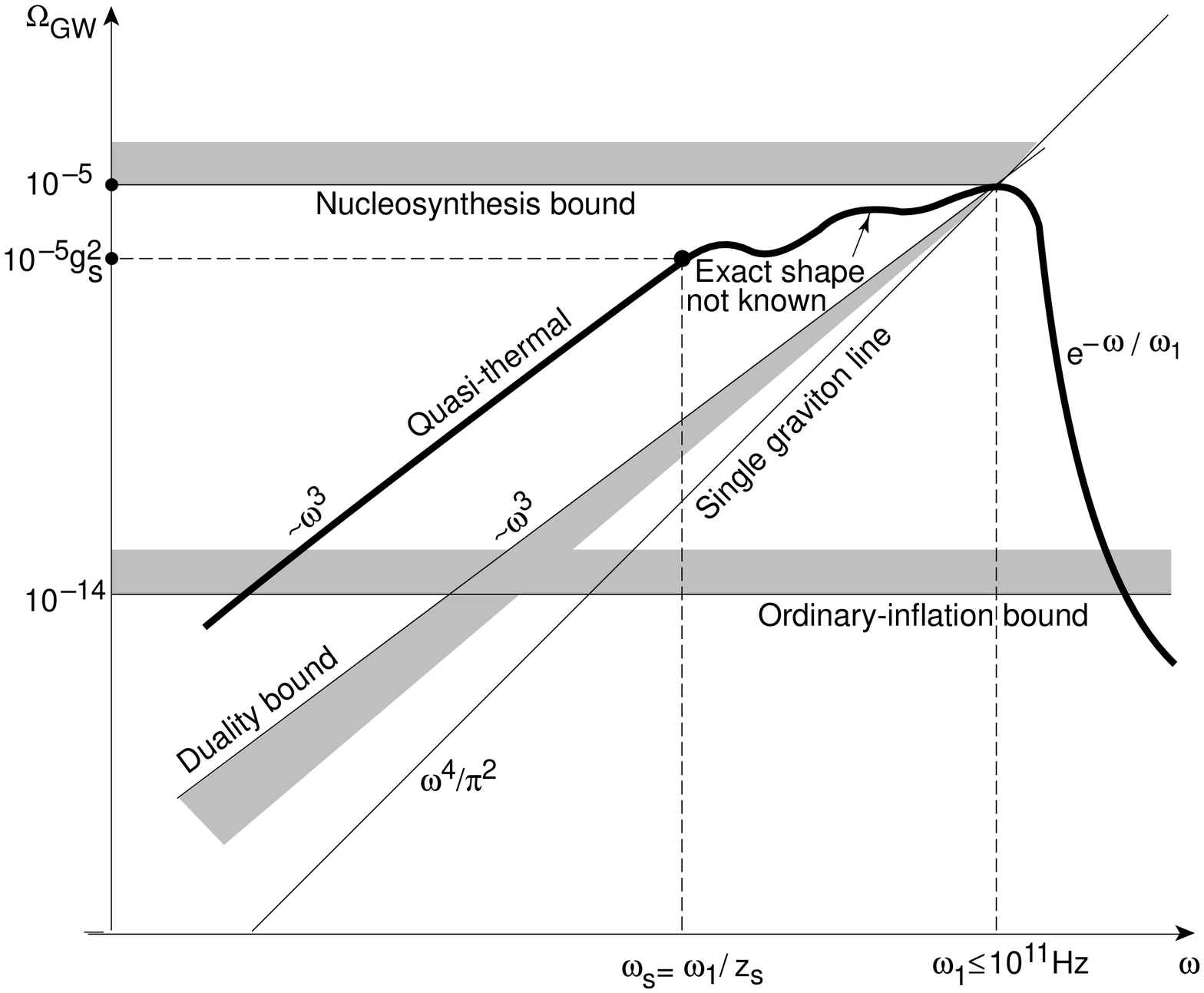,height=3in}
\begin{center} Figure 5 \end{center}
\end{figure}

The final outcome for the GW spectrum \cite{BGGV}, \cite{BGV}
 is shown schematically in Fig. 5 
(leaving a more detailed description to Ramy Brustein's talk).
 For a given pair
$g_s, z_s$ one identifies a point in the $\omega, \Omega_{\omega}$ plane
as illustrated explicitly in the case of $g_s=10^{-3}, z_s= 10^6$. The 
resulting point (indicated by a large dot)
represents the end-point ($\omega_s, \Omega_{\omega_s}$)
of the $\omega^{3}$ spectrum corresponding to scales
having crossed the horizon during the dilatonic era.

Although the rest of the spectrum is more uncertain, it can be argued\cite{BGV}
that it must smoothly  join the point ($\omega_s, \Omega_{\omega_s}$) 
to the true 
end-point $\Omega \le 10^{-5}, \omega \sim 10^{11}$ Hz. The latter
corresponds  to a few gravitons produced at the maximal
amplified frequency $\omega_1$, the last scale to go outside the horizon during
the stringy phase.
The full spectrum  
is also shown in the figure for the  case  $g_s=10^{-3}, z_s=10^6$, with the
wiggly line representing the less well-known high-frequency part.

 If $g_s <1$, as we have
assumed, spectra will always lie below the $ \Omega_{GW}= 10^{-5}$, a line 
representing also a phenomenological bound for a successful
nucleosynthesis to take place \cite{NSB}. On the other
hand, by invoking duality properties of the GW spectrum \cite{duality}, it can
be argued that the actual spectrum will
never lie below the self-dual  spectrum  ending at  $\Omega \sim
10^{-5}, \omega \sim 10^{11}$ Hz (the thick line bordering the shaded region).
In conclusion all possible spectra sweep the angular wedge inside the two
above-mentioned lines and a signal close to the NS bound is all but excluded.
The large  signal can be attributed to the fact that, in the
pre-big-bang scenario, curvatures close to Planck's scale are reached
before the end of inflation.

Having left to the next talk the discussion of further details on the GW
spectrum and on the future prospects of detecting them, I will use the
remaining time to mention a few more encouraging consequences of the
pre-big-bang scenario. Like the generation of GW, they have
something to do with the well-known phenomenon \cite{vv} of
amplification of vacuum quantum fluctuations in
cosmological backgrounds.

The first concerns scalar perturbations: Do they remain small enough during the
pre-big-bang not to destroy the quasi-homogeneity of the Universe?
The answer to this question turns out to be yes! This
is not a priori evident since,
in commonly used gauges (see e.g. \cite{longitudinal}) for
scalar perturbations of the metric (e.g. the so-called longitudinal
gauge in which the metric remains diagonal),
 such perturbations appear
to grow very large during the inflationary phase and to destroy
homogeneity or, at least, to prevent the use of linear perturbation
theory.
 In ref. \cite {BGGMV} it was shown that, by a suitable choice of gauge
(an ``off-diagonal" gauge), the growing mode of the perturbation
can be tamed. This can be double-checked by using the so-called
gauge-invariant variables of Bruni and Ellis \cite{BE}.
The bottom line is that scalar perturbations in string cosmology
behave no worse than tensor perturbations.
An interesting question arises here, in connection with the detectability
of scalar perturbations of this type by using spherical antennas.

The second point that I wish to mention concerns a rather unique prediction
of our scenario: the amplification of EM perturbations.
Because of the scale-invariant coupling
of gauge fields in four dimensions, electromagnetic (EM)
perturbations are $\it{not}$ amplified in a conformally flat
cosmological background (even if inflationary).
In string cosmology, the presence of a time-dependent dilaton
in front of the gauge-field kinetic term allows the amplification of EM
perturbations.
Seeds for generating
the galactic magnetic fields through the so-called cosmic-dynamo
mechanism \cite{Zeld} can thus be obtained.

The final outcome can be expressed \cite {GGV}, \cite {LL} in terms of the
 fraction of electromagnetic
energy stored in a unit of logarithmic interval of $\omega$
normalized to the one in the CMB, $\rho_{\gamma}$. One finds:
\beq
r(\omega)=\frac{\omega}{\rho_{\gamma}}
 \frac{d\rho_{B}}{d\omega} \simeq
\frac{\omega ^{4}}{\rho_{\gamma}} |c_{-}(\omega)|^2 \equiv
\frac{\omega^{4}}{\rho_{\gamma}}(g_{re}/g_{ex})^2~.
\label{r}
\end{equation} 
 where $g_{ex}$ ($g_{re}$) refer to the value of the coupling at exit 
(re-entry) of the scale $\omega$ under consideration.

In terms of $r(\omega)$ the condition for seeding the galactic magnetic field 
through ordinary mechanisms of plasma physics is \cite{TW}
\beq
r(\omega_{G})\geq 10^{-34} \; ,
\label{condition}
\eeq
where  $\omega_{G}\simeq (1$
Mpc$)^{-1}\simeq 10^{-14}$ Hz is the galactic scale.
Using the known value of $\rho_{\gamma}$, we thus find, from (\ref{r}) and
(\ref{condition}):
\begin{equation}
 g_{ex} < 10^{-33}~,
\label{rr}
\end{equation}
i.e. a very tiny coupling at the time of exit of the galactic scale.
 
The conclusion is that string cosmology
 stands a unique chance to explain the origin of
the galactic magnetic fields. Indeed, if the seeds of the magnetic fields 
are to be attributed to the amplification of vacuum fluctuations, 
their present magnitude can be interpreted as prime evidence that the
fine structure constant has evolved to its present value
from a tiny one during inflation.
The fact that the needed variation of the coupling constant ($\sim 10^{30}$)
is of the same order as the variation of the scale factor needed to solve
the standard cosmological problems,
can be seen as further evidence for  scenarios 
in which coupling and scale factor grow roughly at the same rate during
inflation.

 Finally, I would like to mention a more theoretical bonus following from the
pre-big-bang picture: a possible explanation of standard cosmology's
 hot initial state. 
  
The question is: Can one arrive at the
hot big bang of the SCM starting from our
``cold" initial conditions?
The reason why a hot Universe can emerge at the end of our
inflationary epochs (phases I and II)
 goes back to an idea of L. Parker \cite{Pa},
 according to which  amplified quantum
fluctuations can
give origin to the CMB itself if Planckian scales are reached.
 
Rephrasing Parker's idea in our context amounts to solving the
following bootstrap-like condition: At which moment, if any,
 will the energy stored in the perturbations reach
 the critical density? The total energy density $\rho_{qf}$ stored
in the amplified vacuum quantum fluctuations is roughly given by:
\beq
\rho_{qf} \sim N_{eff}~ {M_s^4 \over 4 \pi^2}
\left( a_1 / a \right)^4~,
\label{rhoqf}
\eeq
where $N_{eff}$ is the number of effective (relativistic) species,
which get produced (whose energy density decreases like $a^{-4}$)
and $a_1$ is the scale factor at the (supposed) moment of
branch change.
The critical density (in the same units) is given by:
\beq
\rho_{cr} = e^{-\phi} M_s^2 H^2~.
\label{rhocr}
\eeq
 
At the beginning, with $e^{\phi}\ll 1$, $\rho_{qf} \ll \rho_{cr}$;
 but, in the $(-)$ branch
solution, $\rho_{cr}$ decreases faster than $\rho_{qf}$ so that,
at some moment, $\rho_{qf}$ will become the dominant source
of energy while the dilaton kinetic term will become negligible.
It would be interesting to find out what sort of initial
temperatures for the radiation era will come out of this assumption.


\section{Conclusions}

\begin{itemize}
\item  All cosmological inflationary models lead to the prediction of a
stochastic CGRB that should surround us today very much like
 its electromagnetic
analogue.
\item  Standard inflationary models must unfortunately satisfy the constraint
$\Omega_{GW} < 10^{-10} \Omega_{\gamma}$ in the interesting frequency range.
\item  Inflationary models, such as those suggested by string theory, in
which the Hubble parameter grows during inflation and eventually reaches values
$O(\lambda_s^{-1}, \ell_P^{-1})$, evade the above constraint and (may)
naturally lead to $\Omega_{GW} \le 0.1 \Omega_{\gamma}$ in the interesting
frequency range. 
\item  Observation of such a CGRB would open a unique window on the very early
Universe and thus on fundamental physics at the Planck (string) scale.
\item  Last but not least, as emphasized to me by Emilio Picasso,
trying to detect a stochastic CGRB  is not just relying
on getting a gift from the sky! 
\end{itemize}


\section*{References}
\begin{thebibliography}{99}
\bibitem{W} S. Weinberg, {\it Gravitation and Cosmology},
John Wiley \& Sons, Inc., New York (1972).
\bibitem{INFL}
 L.F. Abbott and So-Young Pi (eds.),
{\it Inflationary Cosmology},
 World Scientific, Singapore
 (1986); \\
 E. Kolb and M. Turner, {\it The
 Early Universe}, Addison-Wesley, New York (1990).
\bibitem{CO} G. Smoot et al., {\it Astrophys. J.} {\bf 396} (1992) L1.
\bibitem{VE}  G. Veneziano,
``Quantum strings and the constants of Nature", in
{\it The Challenging Questions}, Erice, 1989, ed. A. Zichichi, Plenum
Press, New York (1990).
\bibitem{EFF} C. Lovelace, \PL {\bf B135} (1984) 75;\\
 C.G. Callan, D. Friedan, E.J. Martinec and M.J. Perry,
\NP {\bf B262} (1985) 593; \\
 E.S. Fradkin and A.A. Tseytlin,
 \NP {\bf B261} (1985) 1.
\bibitem{JBD} P. Jordan, {\it Z. Phys.} {\bf 157} (1959) 112; \\
C. Brans and R.H. Dicke, \PR {\bf 124} (1961) 925.
\bibitem{EP} See, for instance, E. Fischbach and C. Talmadge,
{\it Nature} {\bf 356} (1992) 207.
\bibitem{TV} T.R. Taylor and G. Veneziano, \PL {\bf B213} (1988) 459.
\bibitem{Wi} E. Witten, \PL {\bf B149} (1984) 351.
\bibitem{SV}  N. Sanchez and G. Veneziano, 
 \NP {\bf B333} (1990) 253;
\bibitem{SFD} G. Veneziano, \PL {\bf B265} (1991) 287.
\bibitem{PBB} M. Gasperini and G. Veneziano,
{\it Astropart. Phys.} {\bf 1} (1993) 317; \MPL {\bf A8} (1993) 3701;
\PR {\bf D50} (1994) 2519.
\bibitem{rr} A.A. Tseytlin, \MPL {\bf A6} (1991) 1721;\\
A.A. Tseytlin and C. Vafa, \NP {\bf B372} (1992) 443.
\bibitem{KK} E. Kiritsis and C. Kounnas, \PL {\bf B331} (1994) 51; \\
A.A. Tseytlin, \PL {\bf B334} (1994) 315.
\bibitem{top} P. Aspinwall, B. Greene and D. Morrison, \PL
{\bf B303} (1993) 249;\\
E. Witten, \NP {\bf B403} (1993) 159; \\
A. Strominger, {\it Massless black holes and conifolds in string theory},
hep-th/9504090.
 \bibitem{BV} R. Brustein and G. Veneziano, \PL {\bf B329} (1994) 429.
 \bibitem{KMO} N. Kaloper, R. Madden and K. A. Olive,
{\it Towards a singularity-free inflationary universe?},
Univ. Minnesota preprint
UMN-TH-1333/95 (June 1995).
 \bibitem {BGGV} R. Brustein, M. Gasperini, M. Giovannini
 and G. Veneziano,  \PL {\bf B361} (1995) 45;\\
 see also M. Gasperini and M. Giovannini, \PR {\bf D47} (1992) 1529.
\bibitem {BGV} R. Brustein, M. Gasperini and G. Veneziano,
{\it Peak and end point of the relic graviton background in string cosmology},
hep-th/9604084; \\
A. Buonanno, M. Maggiore and C. Ungarelli, 
{\it Spectrum of relic gravitational waves in string cosmology}, Pisa preprint 
IFUP-TH 25/96.
\bibitem {NSB} N. Hata et al., \PRL {\bf 75} (1995) 3977; \\
C. Copi et al., \PRL {\bf 75} (1995) 3981; and references therein.
\bibitem{duality} R. Brustein, M. Gasperini and G. Veneziano, in preparation.
\bibitem{vv} L.P. Grishchuk, {\it Sov. Phys. JEPT} {\bf 40} (1975)
409;\\
A.A. Starobinski, {\it JEPT Lett.} {\bf 30} (1979) 682;\\
V.A. Rubakov, M. Sazhin and A. Veryaskin, \PL {\bf B115} (1982) 189;\\
R. Fabbri and M. Pollock, \PL {\bf B125} (1983) 445.
\bibitem{longitudinal} See, for instance,
V. Mukhanov, H.A. Feldman and R. Brandenberger,
{\it Phys. Rep.} {\bf 215} (1992) 203.
\bibitem {BGGMV} R. Brustein, M. Gasperini, M. Giovannini,
V. Mukhanov and G. Veneziano, \PR {\bf D51} (1995) 6744.
\bibitem{BE} G. F. R. Ellis and M. Bruni,
\PR {\bf D40} (1989) 1804;\\ M. Bruni,  G. F. R. Ellis and
P. K. S. Dunsby, {\it Class. Quant. Grav.} {\bf 9} (1992) 921.
\bibitem{Zeld} E. N. Parker, {\it Cosmical Magnetic Fields},
Clarendon, Oxford (1979); \\
 Y. B. Zeldovich, A. A. Ruzmaikin and D. D. Sokoloff,
{\it Magnetic fields in astrophysics},
Gordon and Breach, New York (1983).
\bibitem {GGV} M. Gasperini, M. Giovannini
 and G. Veneziano,  \PRL {\bf 75} (1995) 3796; \PR {\bf D52} (1995) 6651.
\bibitem{LL} D. Lemoine and M. Lemoine, {\it Primordial magnetic fields
in string cosmology}, Inst. d'Astrophysique de Paris preprint (April
1995).
\bibitem{TW} M. S. Turner and L. M. Widrow,
\PR {\bf D37}  (1988) 2743.
\bibitem{Pa} L. Parker, {\it Nature} {\bf 261} (1976) 20.

\end{thebibliography}
\end{document}